\documentclass[11pt,a4paper]{article}
\pdfoutput=1
\usepackage{jheppub}
\usepackage{subfig}

\newcommand{\del}{\ensuremath{\partial}}
\newcommand{\ba}{\begin{eqnarray}}
\newcommand{\ea}{\end{eqnarray}}
\newcommand{\Tr}{\textrm{Tr}}

\title{Ensemble fermions for electroweak dynamics and the fermion preheating temperature.}

\author[a]{Zong-Gang Mou,}
\author[a]{Paul M. Saffin,}
\author[b]{Anders Tranberg}

\affiliation[a]{School of Physics and Astronomy, University Park, University of Nottingham,\\ Nottingham NG7 2RD, United Kingdom}
\affiliation[b]{Faculty of Science and Technology, University of Stavanger, \\4036 Stavanger, Norway}

\emailAdd{ppxzm1@nottingham.ac.uk}
\emailAdd{paul.saffin@nottingham.ac.uk}
\emailAdd{anders.tranberg@uis.no}

\abstract{We refine the implementation of ensemble fermions for the electroweak sector of the Standard Model, introduced in \cite{us2}. We consider the behavior of different observables as the size of the ensemble is increased and show that the dynamics converges for ensemble sizes small enough that simulations of the entire electroweak sector become numerically tractable. We apply the method to the computation of the effective preheating temperature during a fast electroweak transition, relevant for Cold Electroweak Baryogenesis. We find that this temperature is never below 20 GeV, and this in combination with the results of \cite{CPV1} convincingly rules out Standard Model CP-violation as the origin of the baryon asymmetry of the Universe.}


\begin{document}

\maketitle

%%%%%%%%%%%%%%%%%%%%%%%%%%%%%%%%%%%%%%%%%%%%%%%%%%%%%
\section{Introduction}
\label{sec:intro}
%%%%%%%%%%%%%%%%%%%%%%%%%%%%%%%%%%%%%%%%%%%%%%%%%%%%%

The non-equilibrium dynamics of the electroweak sector of the Standard Model and its extensions is crucial for the understanding of baryogenesis and leptogenesis in the early Universe. A large body of work exists on equilibrium quantities including the sphaleron rate (see \cite{sphaleron} for recent results) and electroweak phase diagram \cite{rummukainen}, based on a dimensionally reduced version of the theory, and out-of-equilibrium dynamics has been studied using the classical approximation for the bosonic degrees of freedom (see for instance \cite{clas1,clas2,clas3,clas4,CEB4,kappadep,quench,konstandin}). 

However, a complete understanding of, in particular, fermion production and the baryon asymmetry requires us to include fermionic degrees of freedom, and these are inherently quantum mechanical. It has been known for some time how to combined quantum fermions with classical bosonic fields out of equilibrium \cite{ferm1,ferm2,ferm3,ferm4,ferm5}, using either the complete set of quantum modes, or a statistical ensemble approach. Recently this method was implemented on a lattice for the reduced Standard Model; including only SU(2) gauge fields, a Higgs field and a single family of mass-degenerate quarks and leptons \cite{us1,us2}.

It was demonstrated that the chiral anomaly is indeed generated in this formalism, and that a sizable number (tens of thousands) of realizations are required for the ensemble fermion number to converge, in particular for large values of the fermions-scalar Yukawa coupling. In the Standard Model, the top quark has a coupling of approximately 1, requiring another factor of ten in ensemble size. Although it was highly encouraging that the anomaly is reproduced, the massive numerical effort threatened to make the method unmanageable for realistic systems. 

In this work, we point out that the slow convergence of the fermion number is irrelevant to the dynamics of the fields, but is a feature of the observable. We refine the analysis and show that all other observables of interest converge at much smaller ensemble size, and in particular that the back reaction of fermions onto the bosonic fields converges faster. In practice, this means that simulations are reliable with of order 2000-3000 fermion realisations, which is certainly numerically tractable.

We also introduce a way to compute the particle spectrum and extract an effective temperature from the fermions, using as a testing ground a fast electroweak quench transition. As a direct application of this, we calculate this temperature as a function of the quench rate. This number enters in recent computations of effective CP-violation in Cold Electroweak Baryogenesis \cite{CEB1,CEB2,CEB3,CEB4,CPV1,CPV2}.

The structure of the paper is as follows: In section \ref{sec:model} we set up the reduced Standard Model and in section \ref{sec:modelling} the ensemble fermion method. In section \ref{sec:global} we investigate the behavior and convergence properties of global variables such as the Chern-Simons number, Higgs winding number, fermion number and the energy. Section \ref{sec:spec} discusses the measurement of the particle number and the effective temperature, and in section \ref{sec:fermT} we perform the full simulations to find the quench time dependence, further averaging over an ensemble of bosonic fields. We conclude in section \ref{sec:conc}.

%%%%%%%%%%%%%%%%%%%%%%%%%%%%%%%%%%%%%%%%%%%%%%%%%%%%%
\section{The reduced Standard Model}
\label{sec:model}
%%%%%%%%%%%%%%%%%%%%%%%%%%%%%%%%%%%%%%%%%%%%%%%%%%%%%

We will consider a simplified Standard Model including the SU(2) gauge field $W_\mu^a$ coupled to the Higgs doublet $\phi$ and one family of fermions; a left-handed quark doublet  $q_L=\left(u_L,d_L\right)$, and two right handed singlets $u_R$, $d_R$, and similarly a left-handed lepton doublet $l_L=\left(\nu_L,e_L\right)$ and two right handed singlets $e_R$, $\nu_R$, including a right-handed neutrino. As a consequence, we ignore hypercharge and gluonic gauge fields as well as the second and third family of fermions. 
The continuum action is then written as
\ba
\label{eq:contAcReduced}
S=S_H+S_W+S_{F}+S_{Y},
\ea
with the components
\ba
S_H&=&-\int\;d^4x\;\left[D_\mu\phi^\dagger D^\mu\phi-\mu^2 \phi^\dagger\phi +\lambda(\phi^\dagger\phi)^2\right],\\
S_W&=&-\int\;d^4x\;\frac{1}{4}W^a_{\mu\nu}W^{a,\mu\nu},\\
S_F&=&-\int\; d^4x\;\left[ \bar{q}_L\gamma^\mu D_\mu q_L
                          +\bar{u}_R\gamma^\mu D_\mu u_R+\bar{d}_R\gamma^\mu D_\mu d_R\right.\nonumber\\
                          &~&\qquad\qquad\left.+\bar{l}_L\gamma^\mu D_\mu l_L
                          +\bar{\nu}_R\gamma^\mu D_\mu \nu_R+\bar{e}_R\gamma^\mu D_\mu e_R\right],\\
S_{Y}&=&-\int\; d^4x\;\left[ 
          G^u\bar{q}_L\phi u_R+G^d\bar{q}_L\phi d_R+G^e\bar{l}_L\phi e_R+G^\nu\bar{l}_L\phi \nu_R\right.\\\nonumber
       &~&\qquad\qquad+\hat G^u\bar{q}_L\tilde\phi u_R+\hat G^d\bar{q}_L\tilde\phi d_R+\hat G^e\bar{l}_L\tilde\phi e_R+\hat G^\nu\bar{l}_L\tilde\phi\nu_R\\\nonumber
       &~&\left.\qquad\qquad+h.c.\right].
\ea
The covariant derivatives are
\ba
\label{eq:covDer_phi}
D_\mu\phi&=&\left(\del_\mu-\frac{ig}{2}\sigma^aW^a_\mu\right)\phi,\\
D_\mu q_L&=&\left(\partial_\mu-\frac{ig}{2}\sigma^aW^a_\mu\right)q_L,\quad D_\mu u_R=\partial_\mu u_R,\quad D_\mu d_R=\partial_\mu d_R,\\
D_\mu l_L&=&\left(\partial_\mu-\frac{ig}{2}\sigma^aW^a_\mu\right)l_L\quad D_\mu e_R=\partial_\mu e_R,\quad D_\mu \nu_R=\partial_\mu \nu_R,
\ea
and the SU(2) field-strength is defined by
\ba
\left[D_\mu,D_\nu\right]\phi&=&-\frac{ig}{2}\sigma^aW^a_{\mu\nu}\phi.
\ea
The Higgs mass is taken to be 125 GeV.
It is convenient to redefine the Fermi fields so the kinetic terms have the standard Dirac form, with vector-like gauge-fermion interactions 
\ba
\Psi_R=\epsilon\mathcal{C}^{-1}\bar l_L^T
&\qquad&
\Psi_L=q_L,\\
\chi_R=u_R,
&\qquad&
\chi_L=\mathcal{C}^{-1}\bar e_R^T\\
\xi_R=d_R,
&\qquad&
\xi_L=\mathcal{C}^{-1}\bar \nu_R^T,
\ea
with $\epsilon=i\sigma^2$, and $\mathcal{C}={\rm diag}(-\epsilon,\epsilon)$. 
It follows that
\ba
\bar l_L\gamma^\mu\del_\mu l_L\equiv\bar\Psi_R\gamma^\mu\del_\mu\Psi_R,&\qquad&
\bar e_R\gamma^\mu\del_\mu e_R\equiv\bar\chi_L\gamma^\mu\del_\mu\chi_L,\\
\bar\nu_R\gamma^\mu\del_\mu\nu_R\equiv\bar\xi_L\gamma^\mu\del_\mu\xi_L,&\qquad&
\bar l_L\gamma^\mu\sigma^a W^a_\mu l_L=\bar\Psi_R\gamma^\mu\sigma^a W^a_\mu\Psi_R,
\ea
leaving us with
\ba
S_F&=&-\int\; d^4x\;\left[ \bar\Psi\gamma^\mu D_\mu \Psi
                          +\bar\chi\gamma^\mu \del_\mu \chi+\bar\xi\gamma^\mu \del_\mu \xi\right],\\
S_Y&=&-\int\; d^4x\;\left[G^d\bar\Psi\phi P_R\xi+G^e\bar\chi\tilde\phi^\dagger P_R\Psi+G^u\bar\Psi\tilde\phi P_R\chi-\hat G^\nu\bar\xi\phi^\dagger P_R\Psi+h.c.\right].
\ea
Whereas we before had two left-handed doublets and four right-handed singlets, these are now collected into one full Dirac doublet and two Dirac singlets. The latter only interact via the Yukawa term. For simplicity, we will restrict ourselves to the case
\ba
G^e=G^u=G^d=-G^\nu=\lambda_{\rm yuk},
\ea
which corresponds to all fermions (quarks, charged lepton, neutrino) having the same mass. The global symmetry $q\rightarrow\exp(i\alpha)q$ and $l\rightarrow\exp(i\tilde\alpha)l$ implies classical conservation of the currents
\ba
j^\mu_{(b)}&=&i\left[\bar q_L\gamma^\mu q_L+\bar u_R\gamma^\mu     u_R+\bar d_R\gamma^\mu d_R\right]=i\bar q\gamma^\mu q,\\
j^\mu_{(l)}&=&i\left[\bar l_L\gamma^\mu l_L+\bar \nu_R\gamma^\mu \nu_R+\bar e_R\gamma^\mu e_R\right]=i\bar l\gamma^\mu l,
\ea
and in terms of the redefined fields, we find that
\ba\nonumber
\left(j^\mu_{(5)}\right)_{\rm C-conjugated}=\left(j^\mu_{(b)}+j^\mu_{(l)}\right)_{\rm Original}&=&i\left[-\bar\Psi\gamma^\mu\gamma^5\Psi+\bar\chi\gamma^\mu\gamma^5\chi+\bar\xi\gamma^\mu\gamma^5\xi\right].\\
\ea
At the quantum level, the baryon and lepton currents are no longer conserved due to the chiral anomaly, and
\ba
\del_\mu j^\mu_{(b)}&=&\del_\mu j^\mu_{(l)}=\frac{n_f}{32\pi^2}\left[\frac{1}{2}\epsilon^{\mu\nu\rho\sigma}W^a_{\mu\nu}W^a_{\rho\sigma}\right],\\
   &=&\del_\mu K^\mu.
\ea
where
\ba
K^\mu&=&\frac{n_f}{16\pi^2}\epsilon^{\mu\nu\rho\sigma}\left[W^a_{\nu\rho}W^a_\sigma-\frac{2}{3}\epsilon_{abc}W^a_\nu W^b_\rho W^c_\sigma\right].
\ea
and $n_f$ is the number of fermion families, here taken to be one. The baryon number, $N_f=\int d^3x j^0_{(b)}$, is related to the Chern-Simons number, $N_{CS}=\int d^3x K^0$, as 
\ba
N_f&=&N_{CS}.
\ea
It was demonstrated in \cite{us1,us2} that this relation is reproduced on a lattice using ensemble fermions as described in the following.

%%%%%%%%%%%%%%%%%%%%%%%%%%%%%%%%%%%%%%%%%%%%%%%%%%%%%
\section{Ensemble fermions}
\label{sec:modelling}
%%%%%%%%%%%%%%%%%%%%%%%%%%%%%%%%%%%%%%%%%%%%%%%%%%%%%

The continuum model above is discretized on a lattice of $V/a^3=N^3$ sites, as described in \cite{us2}. The dynamics of bosonic fields is assumed to be classical, and their time evolution follows from a simple variation of the action. Similarly, since the fermions are bi-linear in the action, variation with respect to the fields leads to linear fermion equations of motion involving the bosonic fields, and these equations are solved simultaneously with the bosonic field equations. We deal with the fermion doublers by including a Wilson term in the spatial directions, and by not initialising the lattice time-like doublers. With a small enough time-step, it then takes a long time before these get excited, longer than the duration of our simulations (see also Appendix \ref{app:wilson}). 

In the bosonic equations of motion, fermions enter through bilinear quantum correlators. These can be computed by expanding the fermions on momentum modes in terms of time-independent creation/annihilation operators \cite{ferm1, ferm2}, these operators in turn encode the initial condition. Alternatively one may replace quantum averages by ensemble averages, 
as discussed in \cite{ferm3,us1}.
Since fermion observables can be extracted from the time-ordered Green function, and the fermion back reaction on the bosons is through the equal-time correlation functions, it will be sufficient to give the fermion back reaction, currents and the energy, and demonstrate that these local two-point functions can be well represented by the method.

The equal-time correlation functions for the fermion can be written as,
\ba
D_{\alpha \beta}(\underline x, \underline x';t)=\langle  {\textrm T}\psi_\alpha(\underline x,t)\bar\psi_\beta(\underline x',t)\rangle=\frac{1}{2}\langle \psi_\alpha(\underline x,t)\bar\psi_\beta(\underline x',t)- \bar\psi_\beta(\underline x',t)\psi_\alpha(\underline x,t) \rangle ,
\ea
and we notice that
\ba
\langle  {\textrm T}\bar\psi(\underline x',t)\psi(\underline x,t)\rangle=-\Tr [D(\underline x,\underline x';t)].
\ea
We may expand the field operator as
\ba
\psi(\underline x,t)=\int\frac{d^3p}{(2\pi)^3}e^{ip.x}\psi(\underline p,t)=\sum_s\int\frac{d^3p}{(2\pi)^3}\frac{1}{2\omega_p}\left[b_s(\underline p)U_s(\underline p)e^{ip.x}+d^\dagger_s(\underline p)V_s(\underline p)e^{-ip.x}\right],\nonumber\\
\label{fieldoperator}
\ea
in terms of the annihilation and creation operators $b$ and $d^\dagger$. The fermion anti-commutation relations correspond to
\ba
\{b_r^\dagger(\underline p),b_s(\underline p')\}&=&(2\pi)^3(2\omega_p)\delta_{rs}\delta(\underline p-\underline p'),\\
\{d_r^\dagger(\underline p),d_s(\underline p')\}&=&(2\pi)^3(2\omega_p)\delta_{rs}\delta(\underline p-\underline p'),
\label{antico}
\ea
so the equal-time correlation function can be written out explicitly in the vacuum 
\ba
D(\underline x,\underline x';t)&=&\sum_s\int\;\frac{d^3p}{(2\pi)^3}\frac{1}{2\omega_p}
              \left[U_s(\underline p)\bar U_s(\underline p)e^{ip.(x-x')}-
              V_s(\underline p)\bar V_s(\underline p)e^{-ip.(x-x')}\right].
\label{eq:biLinearQuant}
\ea
We introduce two ensembles of fermions, M(ale) and F(emale), 
\ba
\label{eq:ferminit}
\psi_{M,F}(\underline x,t)&=&\frac{1}{\sqrt{2}}\sum_s\int\;\frac{d^3p}{(2\pi)^3}\frac{1}{2\omega_p}\left[\xi_s(\underline p)U_s(\underline p)e^{ip.x}
                \pm\eta_s(\underline p)V_s(\underline p)e^{-ip.x}\right],
\ea
and where the exact same random numbers $\xi$, $\zeta$ are used in a given Male and Female pair.
We then require that the variables $\xi$ and $\eta$ satisfy the ensemble average relations\footnote{using the standard  analogue lattice version of these correlators and delta functions.}, 
\ba
\langle\xi_r(\underline p)\xi^\star_s(\underline p')\rangle_e&=&(2\pi)^3(2\omega_p)\delta_{rs}\delta(\underline p-\underline p'),\\
\langle\eta_r(\underline p)\eta^\star_s(\underline p')\rangle_e&=&(2\pi)^3(2\omega_p)\delta_{rs}\delta(\underline p-\underline p'),
\ea
we may calculate
\ba
D(\underline x,\underline x';t) = \frac{1}{2}\langle\psi_M(\underline x)\bar\psi_F(\underline x')+\psi_F(\underline x)\bar\psi_M(\underline x')\rangle_e.
\ea
Generating sets of $\eta_k,\xi_k$ and inserting them into eq. (\ref{eq:ferminit}) provides the initial condition for the fermion ensemble fields, and these are solved in position ({\bf x}) space and averaged over to generate the bilinears at each time-step, which are in turn fed into the bosonic equations of motion. The number of realizations in the fermion ensemble is denoted $N_q$, and we must ensure convergence of the physical observables as $N_q$ is increased. In \cite{us2}, it was found that for small values of $\lambda_{\rm yuk}$, $N_q\simeq 10000$ gave convergence of the fermion number observable $N_f$.

The fermion evolution is unitary, and will conserve inner products for each gender, Male or Female
\ba
\sum_x\langle\psi_G^\dagger(\underline x,t)\psi_G(\underline x,t)\rangle_e =\sum_p\langle\psi_G^\dagger(\underline p,t)\psi_G(\underline p,t)\rangle_e = 2N^3,
\ea
where the field is discretized and normalized by the lattice spacing (see \cite{us2}).
There exist local forms
\ba
\langle\psi_G^\dagger(\underline x,t)\psi_G(\underline x,t)\rangle_e = 
\langle\psi_G^\dagger(\underline p,t)\psi_G(\underline p,t)\rangle_e = 2,
\ea
for arbitrary $x$ and $p$. We will use them to monitor the temporal doubler (see also Appendix \ref{app:wilson}).

%%%%%%%%%%%%%%%%%%%%%%%%%%%%%%%%%%%%%%%%%%%%%%%%%%%%%
\section{Global observables}
\label{sec:global}
%%%%%%%%%%%%%%%%%%%%%%%%%%%%%%%%%%%%%%%%%%%%%%%%%%%%%

We will compute a number of different observables to check the convergence of the simulation and determine $N_{qc}$, the smallest $N_q$ for which the observables can be said to have converged.
%Once lattice parameters are fixed, the convergence might seem to be affected by the initial condition or choices of observables. So a safe 
$N_{qc}$ should be robust to different initial conditions that bring different dynamics.

We define the average Higgs field squared, %to demonstrate the motion of the Higgs field,
\ba
\langle\phi^2\rangle =  \frac{2}{{\rm V}} \int d^3x \frac{\phi^\dagger \phi}{v^2}.
\ea
It is scaled to be unity when the Higgs field is in the broken phase vacuum.

The total energy is conserved up to the order $O(a_t/a)$, with energy density components defined from the lagrangian
\ba
e_H&=&\frac{1}{{\rm V}}\int\;d^3x\;\left[D_0\phi^\dagger D^0\phi+D_i\phi^\dagger D^i\phi-\mu^2 \phi^\dagger\phi +\lambda(\phi^\dagger\phi)^2\right],\\
e_W&=&e_E+e_B=\frac{1}{{\rm V}}\int\;d^3x\;\left[\frac{1}{2}W^a_{0i}W^{a,0i}+\frac{1}{4}W^a_{ij}W^{a,ij}\right],\\
e_F&=&\frac{1}{{\rm V}}\int\;d^3x\;\left[ \bar\Psi\gamma^i D_i \Psi
                        +\bar\chi\gamma^i \del_i \chi+\bar\xi\gamma^i \del_i \xi\right],\\
e_Y&=&\frac{1}{{\rm V}}\int\;d^3x\;\lambda_{\rm yuk}\left[\bar\Psi\phi P_R\xi+\bar\chi\tilde\phi^\dagger P_R\Psi+\bar\Psi\tilde\phi P_R\chi+\bar\xi\phi^\dagger P_R\Psi+h.c.\right],\\
e_W&=&-\frac{1}{{\rm V}}\int\; d^3x\;\frac{r_wa}{2}\left[ \bar\Psi D^i D_i \Psi
                          +\bar\chi D^i \del_i \chi+\bar\xi D^i \del_i \xi\right],\\
e_C&=&\frac{1}{{\rm V}} \int d^3x ~\left[\right.\frac{(Z_3-1)}{4} (2W^a_{0i}W^{a,0i}+W^a_{ij}W^{a,ij} ) \nonumber\\
& &~~~~~~~~~~~~~~~~~~~~ + (Z_{\phi}-1) ( D_0\phi^\dagger D^0\phi+D_i\phi^\dagger D^i\phi) + \delta V(\phi) \left.\right].
\ea
The contribution $e_C$ includes the counterterms needed to formally keep the theory finite (see Appendix \ref{app:wilson}). 
The physical fermion energy is obtained by summing over $e_F$, $e_Y$, $e_W$ and $e_C$, and normalized by subtracting its initial value. Then around a well-defined vacuum, the fermion energy can be interpreted as the total energy of all excited particles.

\begin{figure}
\includegraphics[width=1.0\textwidth]{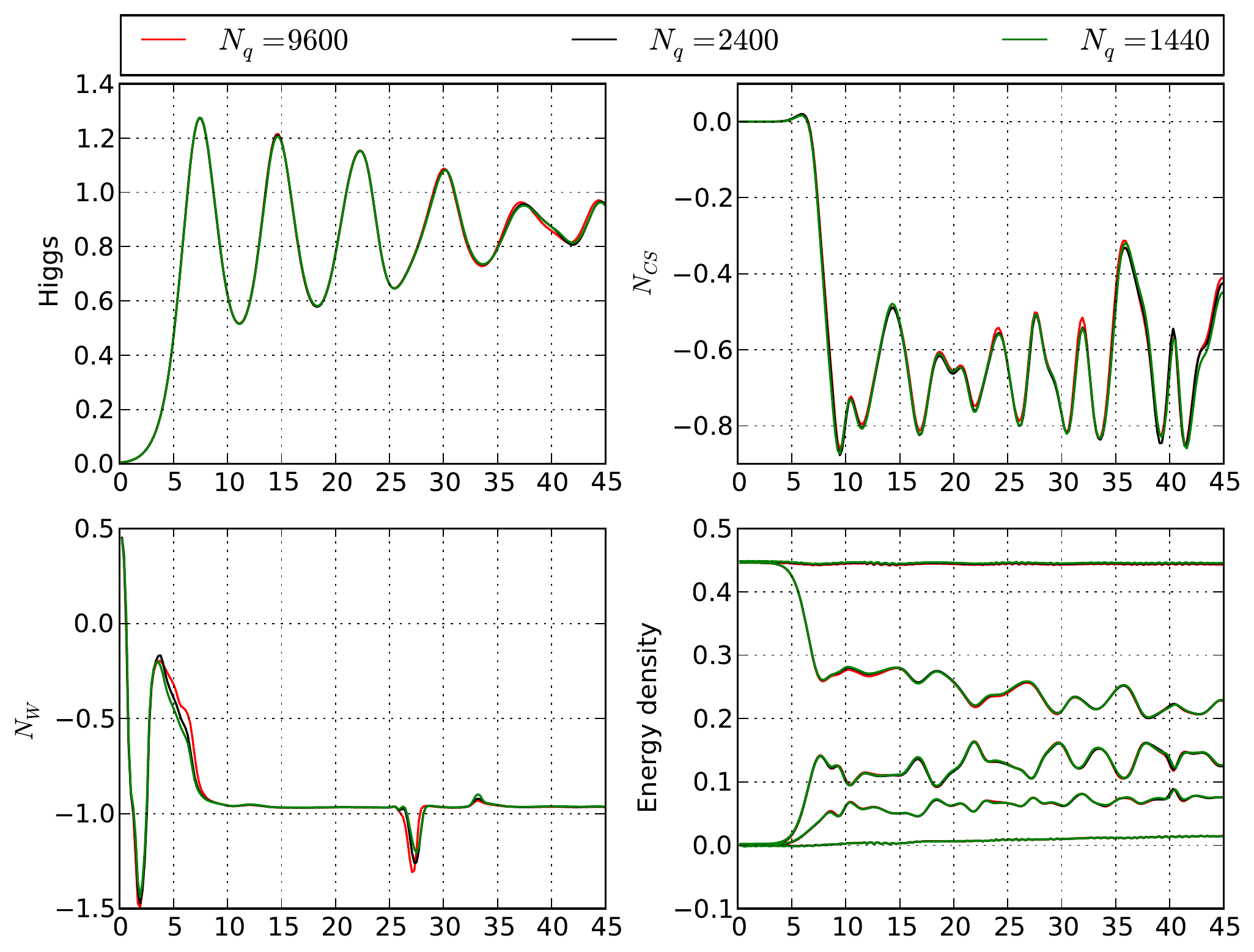}
\caption{ Convergence of Higgs, $N_{CS}$, $N_{W}$ and energy density with $N_q$. $N=32$, $\lambda_{\rm yuk}=0.03$ and $am_H=0.42$. The x-axis is the time.}
\label{fig:nq_higgs_ncs_nw_en}
\end{figure}

We test the dynamics, convergence and observables in a fast electroweak transition, where the Higgs potential undergoes an instantaneous quench $+\mu^2\rightarrow -\mu^2$. In Fig. \ref{fig:nq_higgs_ncs_nw_en}, we show the convergence of a set of global observables (Higgs field, $N_{CS}$, $N_{W}$ and energy). We make a comparison between $N_q=9600$, $2400$ and $1440$, with lattice size $N=32$, Yukawa coupling $\lambda_{\rm yuk}= 0.03$ and the lattice spacing $am_H=0.42$.   We see that convergence is achieved for these observables with $N_q\simeq 1000-2000$ and we take as a conservative choice $N_{qc}=2400$. We found that this was safe also for larger lattices of $N=48$. Convergence is less good for larger Yukawa couplings, but improve for smaller couplings.  We note that $\lambda_{\rm yuk}=0.03$ corresponds to a fermion mass of $5.1$ GeV, so that all Standard Model fermions except the top correspond to smaller values of the coupling.

\begin{figure}
\begin{center}
\includegraphics[width=1.0\textwidth]{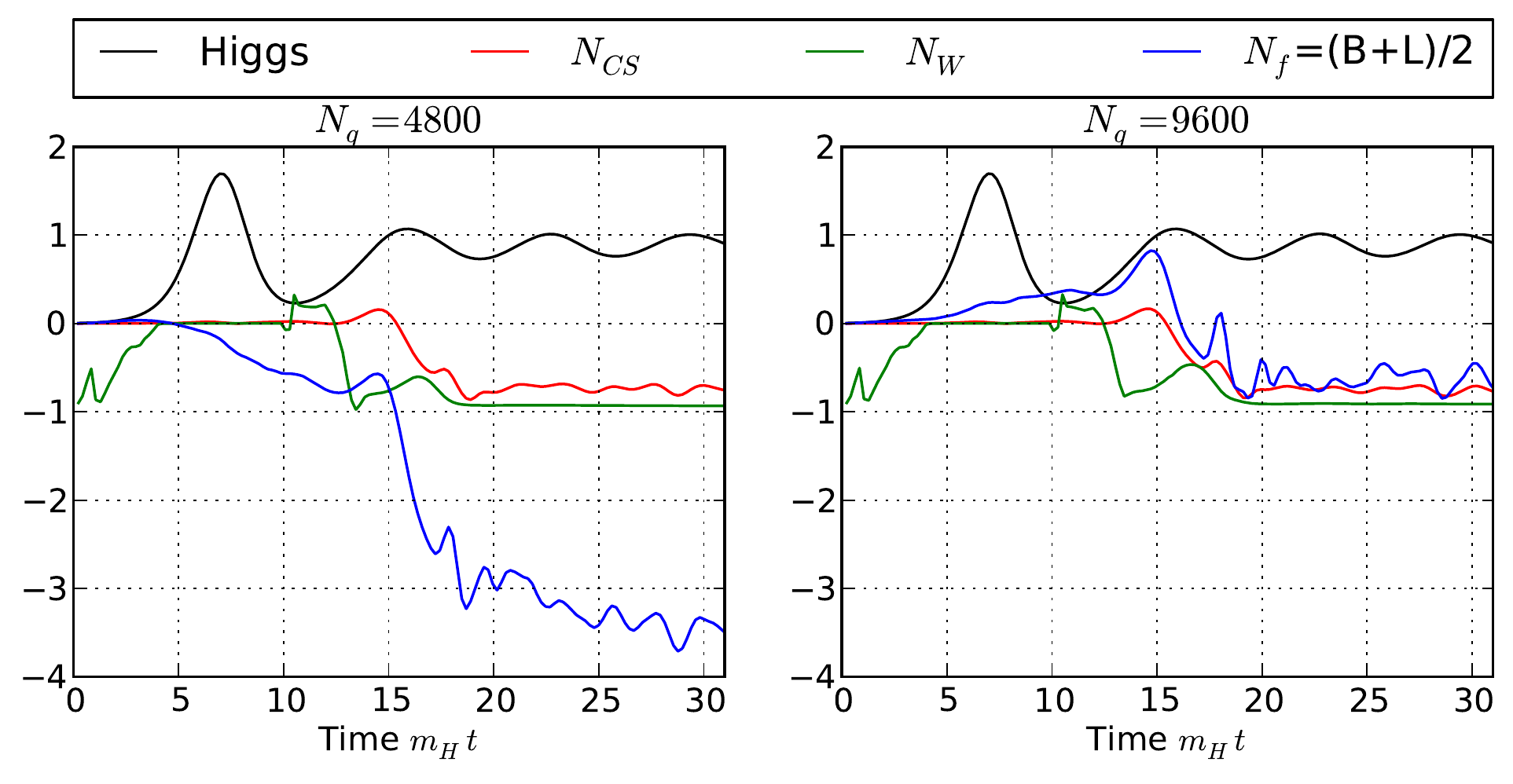}
\caption{The convergence of the baryon (or fermion) number requires much larger $N_q$. }
\label{fig:baryon}
\end{center}
\end{figure}

In Fig. \ref{fig:baryon} we show one special observable, the baryon number (or fermion number), which has not converged yet, even at $N_q=4800$. This is consistent with the findings of  \cite{us2}, but we now see that it is a result of the observable being badly behaved, not the dynamics. In particular, from Fig. \ref{fig:nq_higgs_ncs_nw_en} we can conclude that the fermion back reaction on the bosonic field has also converged at $N_{qc}$, and as a measure of the baryon production, we can simply use bosonic operators, $N_{CS}$ and $N_{W}$. This is with the understanding that had we increased $N_q$ for a factor of 10 or more, $N_f$ would converge to this same number \cite{us2}.

\begin{figure}
\begin{center}
\includegraphics[width=1.0\textwidth]{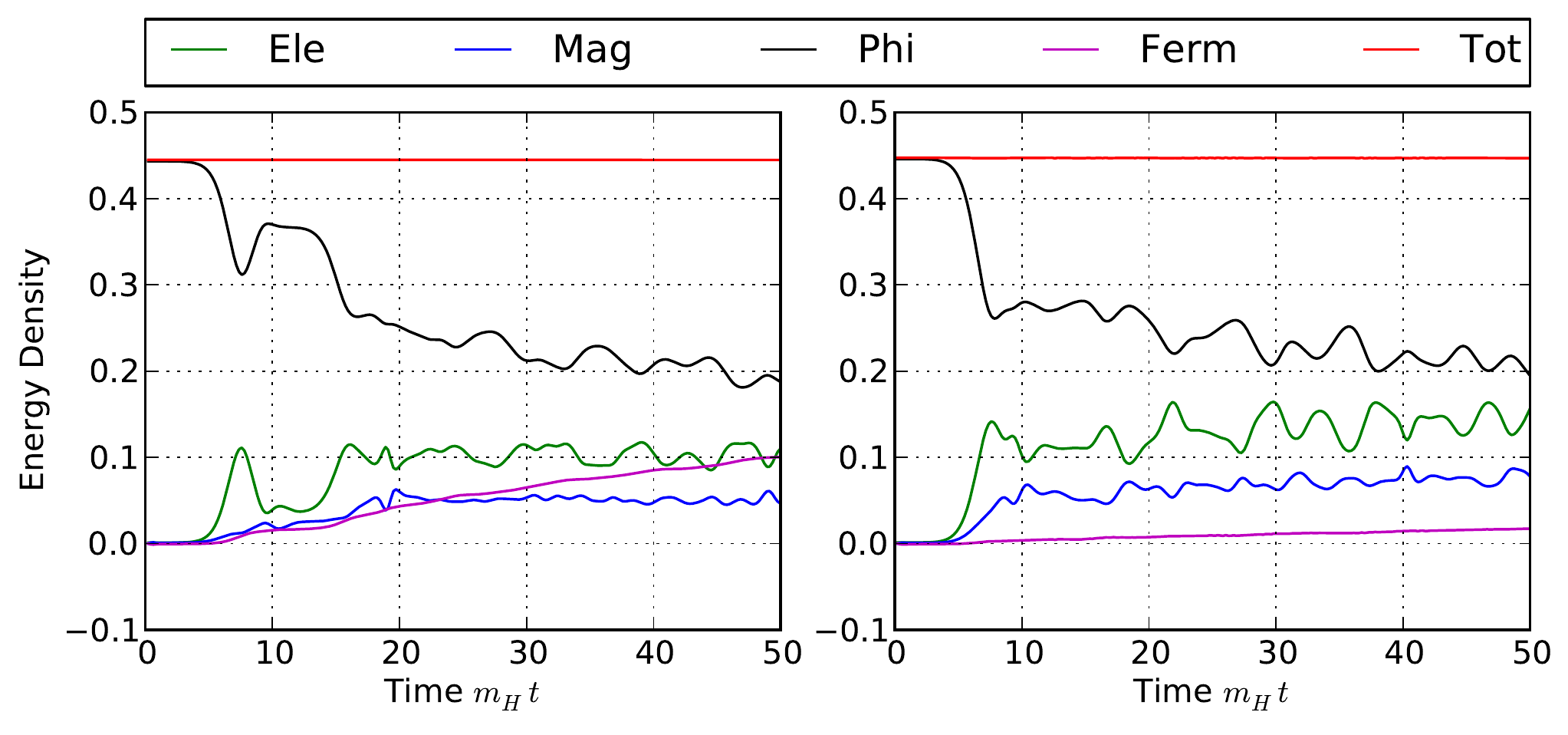}
\caption{Different energy component in a tachyonic electroweak transition. The Wilson coefficient $r_w$ is $0$ in the left plot, and $0.5$ in the right plot. The energy density is scaled by $m^4_H$. More energy is transferred to the fermion when the Wilson term is small.}
\label{fig:rw_en}
\end{center}
\end{figure}

A further example is to look at the individual energy components, shown in Fig. \ref{fig:rw_en} for two different values of the Wilson coefficient $r_w$.
When $r_w=0$ (left plot), the energy transfer to the fermions is much faster than when $r_w=0.5$ (right plot). Apparently the Wilson term is making the doubler modes more heavy, so that they can no longer be excited. And so even when not looking specifically for the effect of the baryon anomaly, it may be worthwhile including the Wilson term in the dynamics. In our simulations, the Wilson parameter $r_w$ is fixed to 0.5.

\begin{figure}
\begin{center}
\includegraphics[width=1.0\textwidth]{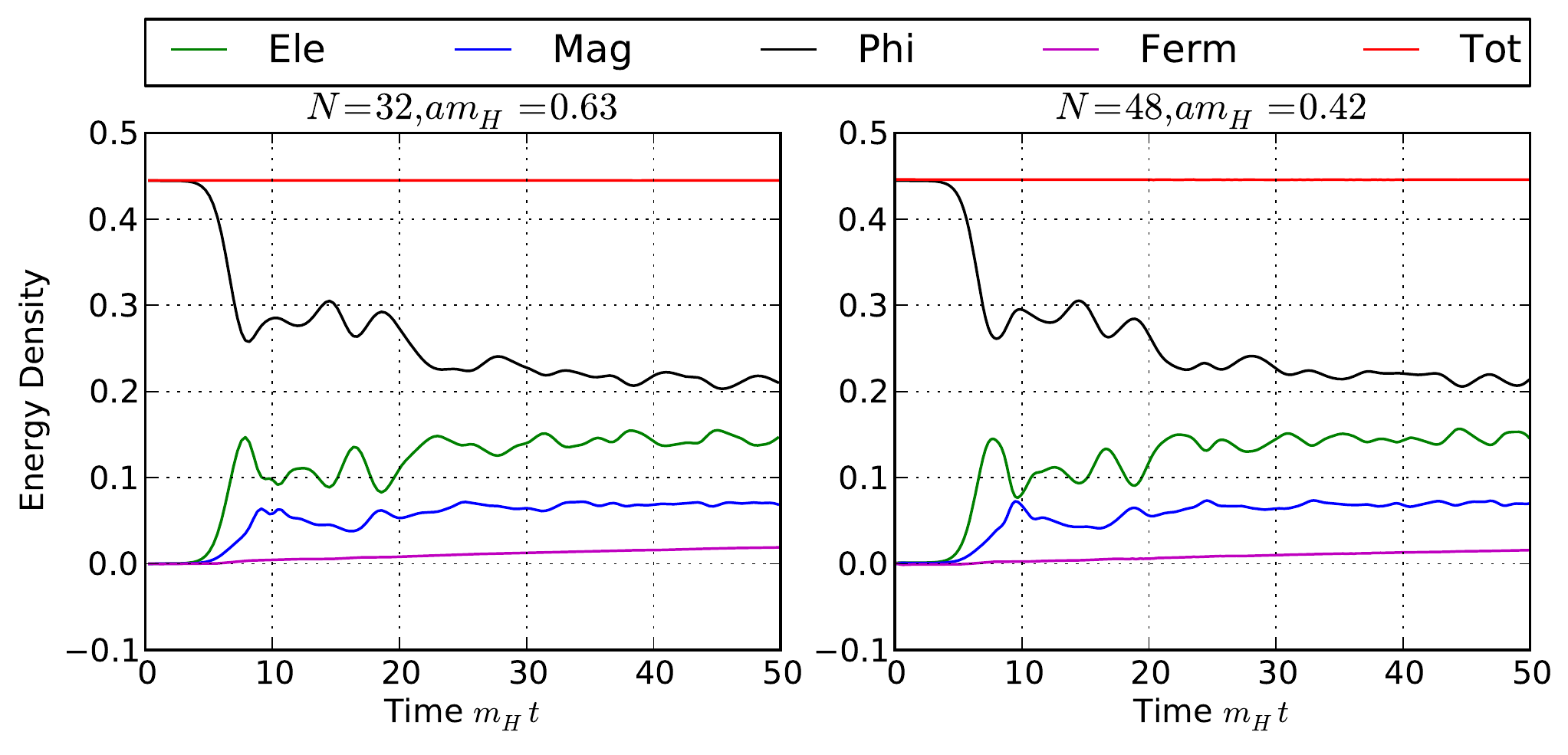}
\caption{The energy components in simulations at different lattice spacing but the same physical volume. Cut-off effects are well under control through tuning the counterterms accordingly. }
\label{fig:a_en}
\end{center}
\end{figure}

The lattice spacing dependence is shown in Fig. \ref{fig:a_en}, showing the different energy components on two different lattices with the same physical volume, but different lattice spacing ($N=32, am_H=0.63$ and $N=48, am_H=0.42$). We see that by choosing the counterterms carefully, the lattice spacing dependence is nicely under control. We note that the random bosonic initial condition is different in the two cases (since there are a different number of modes present), and so the agreement is quite non-trivial.

%%%%%%%%%%%%%%%%%%%%%%%%%%%%%%%%%%%%%%%%%%%%%%%%%%%%%
\section{Spectrum}
\label{sec:spec}
%%%%%%%%%%%%%%%%%%%%%%%%%%%%%%%%%%%%%%%%%%%%%%%%%%%%%%

The fermion particle number can be extracted from the two-point correlation functions. If the fermion field is close to thermal equilibrium, fitting of the particle number with the Fermi-Dirac distribution will give the (effective) temperature and chemical potential of the system.

We define the correlation function in momentum space through a Wigner transform and averaging over the space volume,
\ba
D(\underline p,t) = \frac{1}{{\rm V}}\int d^3X \int d^3z e^{-ip.z}D(\underline X+\frac{1}{2}\underline z,\underline X-\frac{1}{2}\underline z;t),
\ea
which is equivalent to computing $\langle | {\textrm T}\psi(\underline p,t)\bar\psi(\underline p,t)|\rangle$.
For future use, we define:
\ba
F(\underline p,t)  =\Tr[D(\underline p,t)], ~~~~~~~~~~~ 
V(\underline p,t)  =\Tr[i\frac{\underline{\gamma} .\underline{p}}{p}D(\underline p,t)].
\ea

A gauge-invariant correlation function for the gauge doublet can be defined as
\ba
\label{eq:doublet}
D_{\Psi}(x,y;t) = \langle 0| {\rm T} \Psi(x,t)\bar\Psi(y,t) U(y,x;t)|0\rangle,
\ea
where $U(y,x;t)$ is the gauge link connecting  $x$ and $y$ at the time $t$. For the gauge singlet field, this is not needed
\ba
\label{eq:singlet}
D_{\xi,\chi}(x,y;t) = \langle 0| {\rm T} \xi(x,t)\bar\xi(y,t)|0\rangle,\qquad  \langle 0| {\rm T} \chi(x,t)\bar\chi(y,t)|0\rangle.
\ea
Including the gauge link in the definition for the doublet soon becomes cumbersome, and it is not straightforward when rotating to the mass eigenstates composed of linear combinations of singlet and doublet modes. So in practice, we will not consider the gauge invariant correlation functions, but instead fix to unitary gauge. We will, however consider both the weak eigenmodes (where the singlet correlator is gauge invariant) and the mass eigenstates for comparison.

\begin{figure}[ht]
\begin{center}
\includegraphics[width=1.0\textwidth]{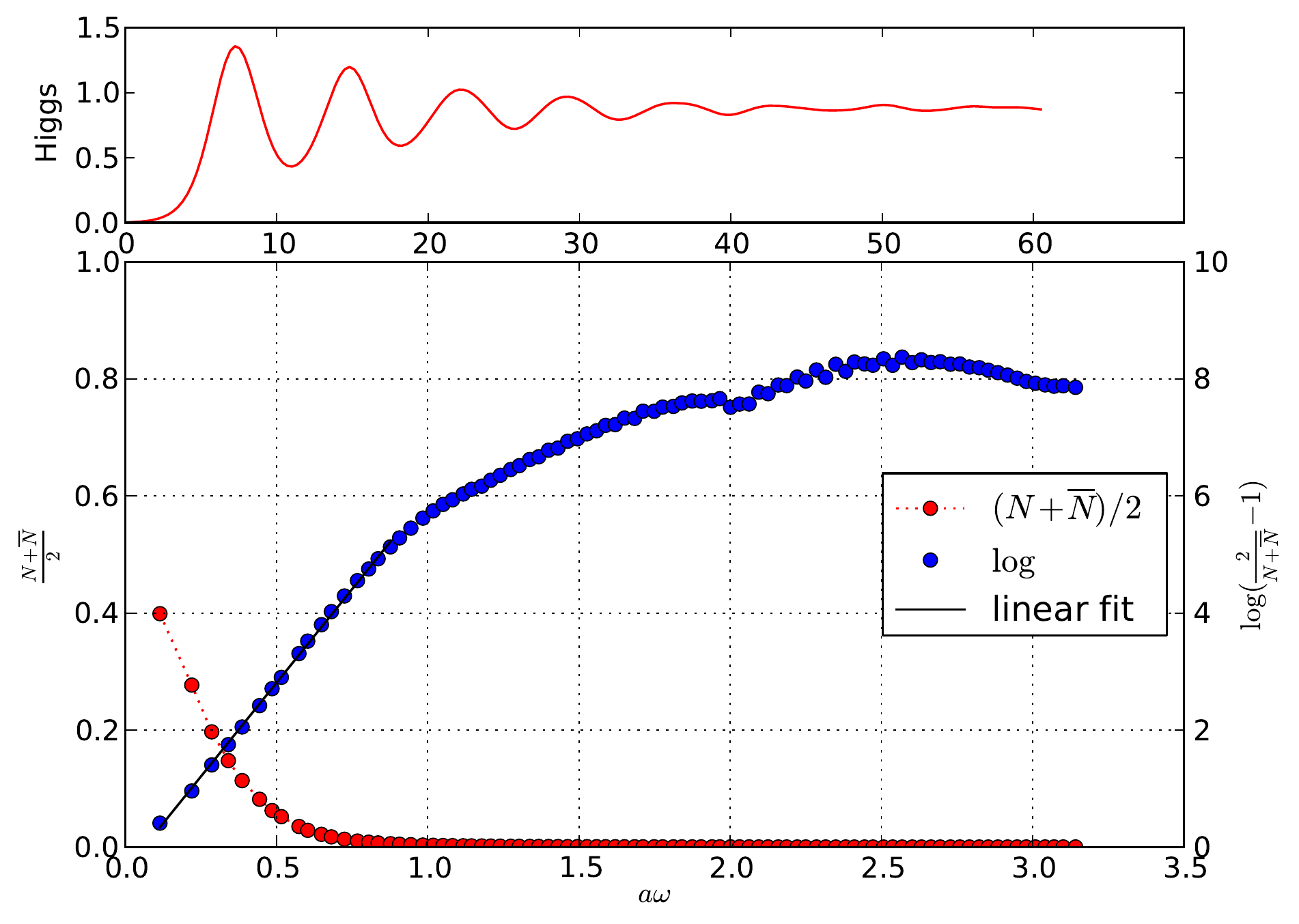} 
\caption{The fermion particle spectrum (bottom panel). The red line is the average particle number, the blue line the derived $\log(1/N_{\rm av}-1)$, so that the slope is the inverse temperature. The black line is the linear fit of the lower energy range. For the linear fitting, we select the range $a\omega\in[0, 0.9]$. The upper plot shows the time evolution of the Higgs field observable and the time ($m_Ht=60.795$) at which the spectrum is computed. $N=48$, $am_H=0.63$ and $\lambda_{\rm yuk}=0.03$.}
\label{fig:spec}
\end{center}
\end{figure}

The correlation function is expected to have the form \cite{ferm1}
\ba
D(\underline p,t)&=& [1-N(\underline p,t)-\bar N(-\underline p,t)] \frac{m(\underline p,t) - i p.\gamma}{2\omega(\underline p,t)} + [\bar N(-\underline p,t)-N(\underline p,t)]\frac{i\gamma^0}{2}, 
\ea
which is obviously true for the setup (\ref{fieldoperator}) and (\ref{antico}).
By assuming the on-shell condition $\omega(p,t)= \sqrt{p^2+m^2(\underline p,t)}$, we have enough freedom to measure the effective energy $\omega(\underline p,t)$ and the average particle number $N_{\rm av}(\underline p,t)\equiv(N(\underline p,t) +\bar N(-\underline p,t))/2$ simultaneously. 
\ba
N_{\rm av}(\underline p,t) &=& \frac{1}{2}-\textrm{sign}[V(\underline p,t)]\frac{\sqrt{F^2(\underline p,t)+ V^2(\underline p,t)}}{4} ,  \\ 
\omega(\underline p,t) &=& \sqrt{\frac{4(1-2N_{\rm av})^2 p^2}{V^2(\underline p,t)}} .
\ea
In the unitary gauge, the mass eigenstates $\Omega_{1,2}$ can be written in terms of the weak eigenstates as
\ba
\label{eq:unitary}
\Omega_1 = \frac{1}{\sqrt{2}} (\Psi_u + \xi),~~~~~
& &\Omega_2 = \frac{1}{\sqrt{2}} (\Psi_d + \chi),
\\
\Omega'_1 = \frac{1}{\sqrt{2}} (\Psi_u - \xi),~~~~~
& &\Omega'_2 = \frac{1}{\sqrt{2}} (\Psi_d - \chi),
\ea  
where $\Psi_u$, $\Psi_d$ are the up and down parts of the gauge doublet.

Having computed the particle number $N_{\rm av}(\omega)$, we can proceed to extract the effective temperature and chemical potential, by fitting to the form
\ba
\log\left(\frac{1}{N_{\rm av}}-1\right) = \frac{\omega-\mu}{T}.
 \ea
 If the system is properly equilibrated, the low-momentum range should indeed show this form (Fig. \ref{fig:spec}).
 
\begin{figure}
\begin{center}
\includegraphics[width=1.0\textwidth]{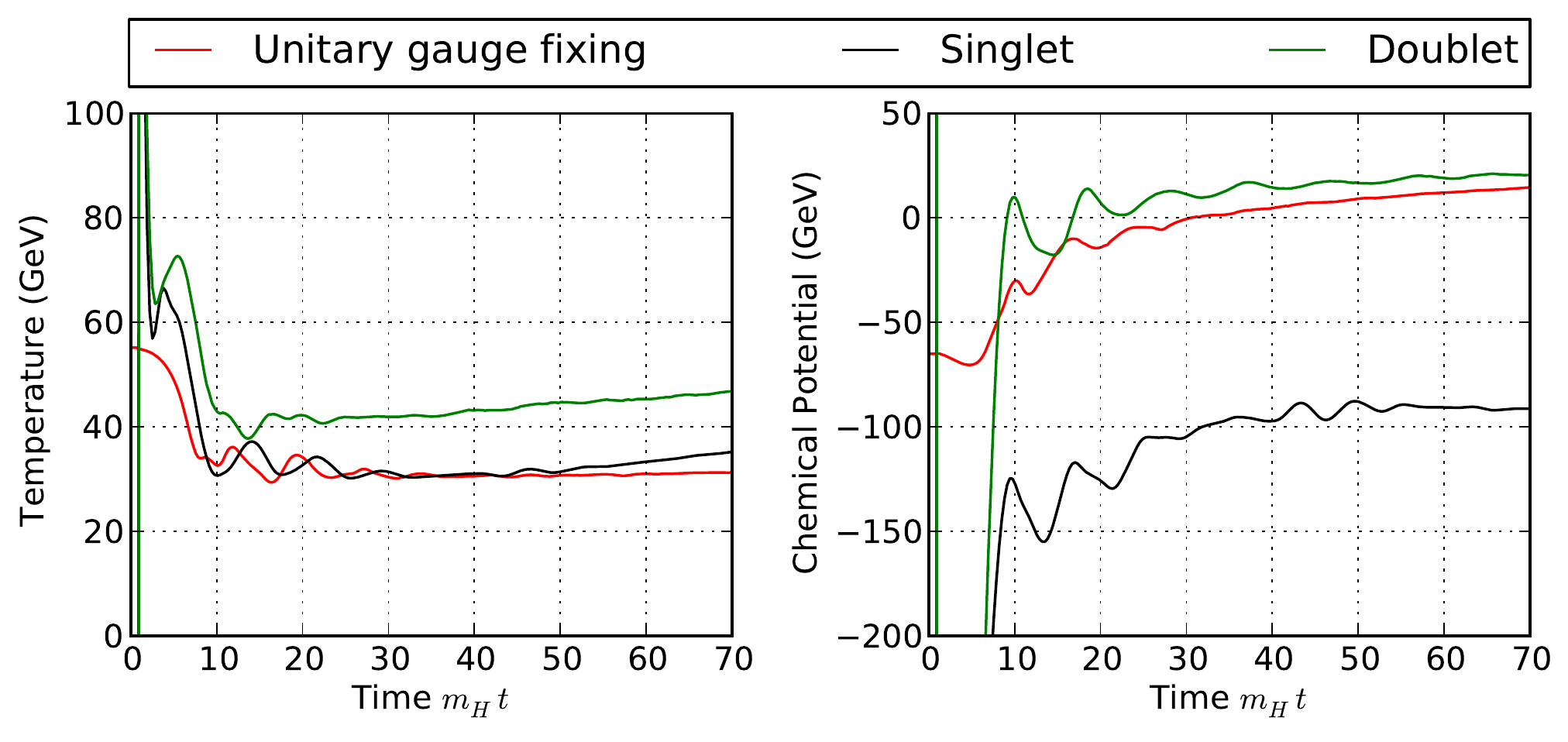}
\caption{ Three definitions of the temperature and chemical potential. The red line is extracted from the mass eigenstate ($\Omega_1$) after the unitary gauge fixing. The singlet and doublet are extracted from the gauge singlet ($\xi$) and gauge doublet ($\Psi_u$) directly. $N=48$, $N_q=2400$, $\lambda_{\rm yuk}$=0.03 and $am_H=0.63$. }
\label{fig:sin_dou}
\end{center}
\end{figure}

To estimate the uncertainty from the gauge non-invariance, we have compared results from different choices of correlators in Fig. \ref{fig:sin_dou} (left). The singlet (field $\xi$) is simply the correlator (\ref{eq:singlet}) and doublet (field $\Psi_u$) is (\ref{eq:doublet}), but omitting the link variable. We compare these to the temperature of an mass eigenstate ($\Omega_1$) in the unitary gauge (\ref{eq:unitary}). We see that the doublet temperature deviates from the other two, with the unitary gauge one the smoothest. The right-hand plot shows the effective chemical potential, which only at later times becomes positive. 
%Again there is good agreement between the singlet and gauge fixed results. 

%Notice the correlation function of the gauge singlet is free of the gauge link, even though the field $\chi$, $\xi$ are not mass eigenstates.
%The temperature of the singlet would be consistent with the temperature of the mass eigenstates measured in the unitary gauge fixing, from time=18, which is just after the %Higgs second rolling-down.
%In practice (Fig. \ref{avspec}), the fitting range should be smaller than the inverse of the lattice spacing.

\begin{figure}
\begin{center}
\includegraphics[width=1.0\textwidth]{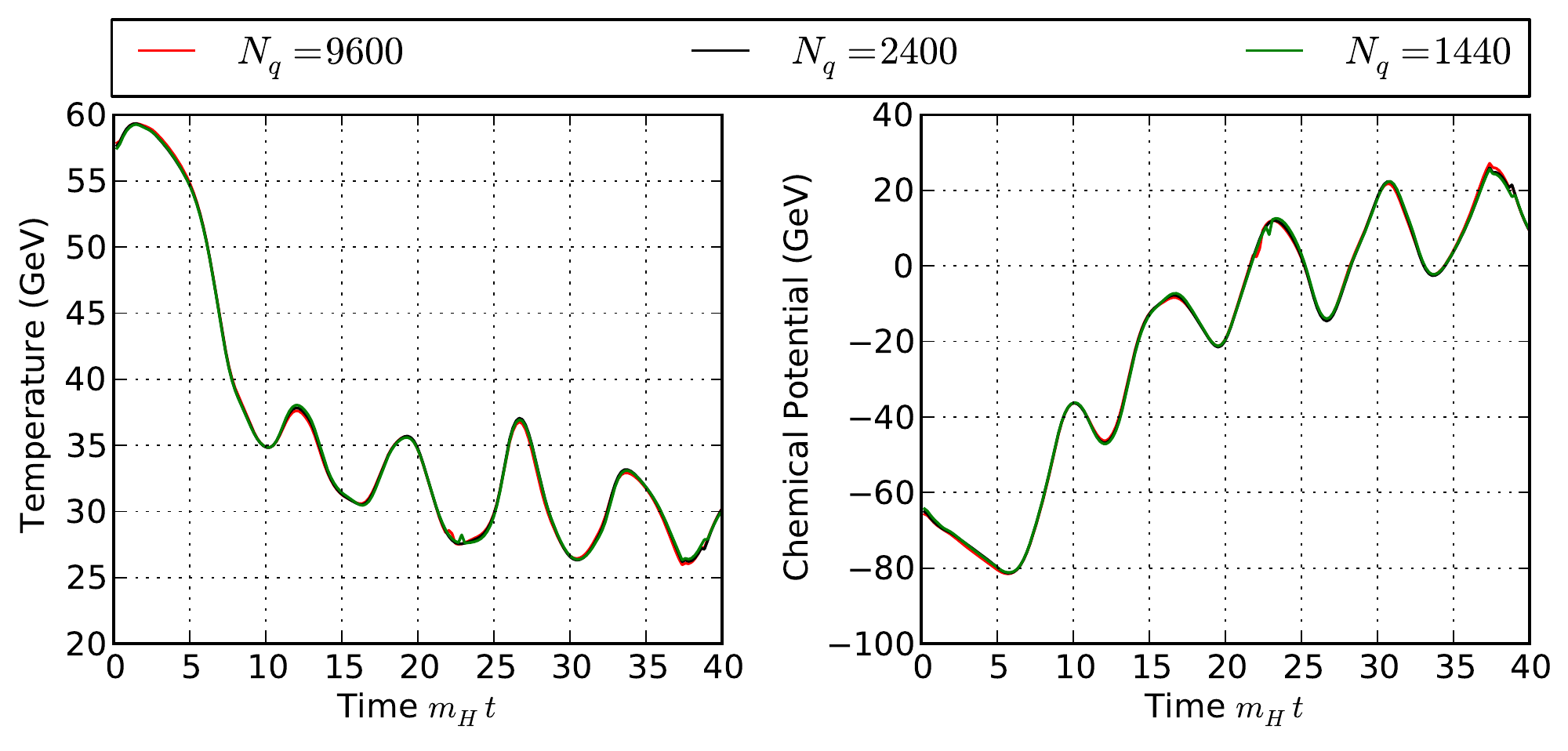}
\caption{ Convergence of the temperature and chemical potential with $N_q$. $N=32$, $\lambda_{\rm yuk}=0.03$ and $am_H=0.42$.}
\label{fig:nq_tem_che}
\end{center}
\end{figure}

\begin{figure}
\begin{center}
\includegraphics[width=1.0\textwidth]{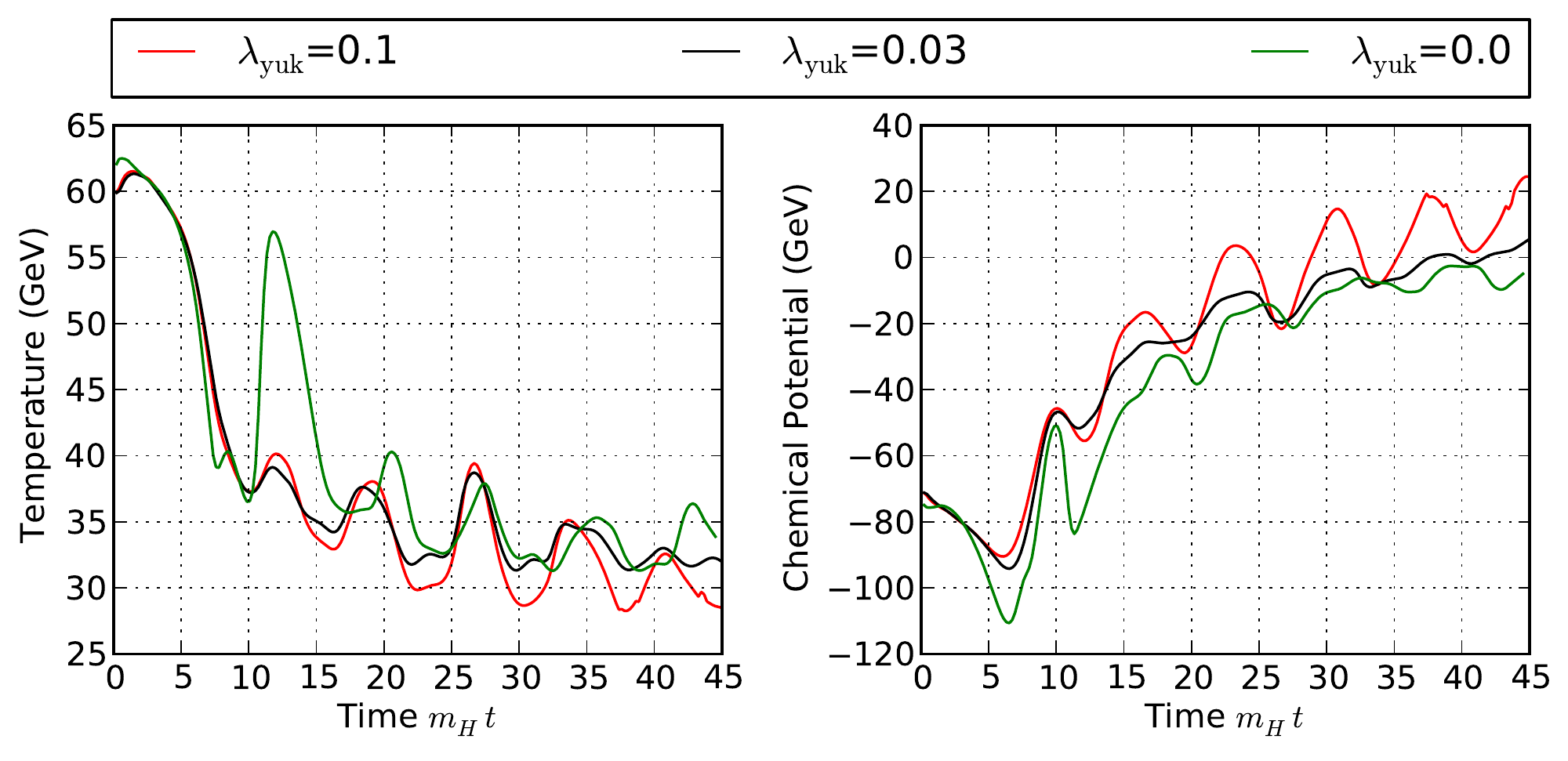}
\caption{The effective temperature and chemical potential for different Yukawa couplings $\lambda_{\rm yuk}=0.1$, $0.03$ and $0$, correspond to fermion masses of $m_f= 17$, $5$ and $0$ GeV respectively. $N_q=2400$, $N=32$, and $am_H=0.42$.}
\label{fig:yuk}
\end{center}
\end{figure}

In Fig. \ref{fig:nq_tem_che} we show a comparison between difference ensemble sizes $N_q$ demonstrating that our choice of $N_{qc}=2400$ also holds for these derived observables. We can also compare results for different values of the Yukawa coupling, Fig. \ref{fig:yuk}.
%Different Yukawa couplings are studied in the Fig. \ref{fig:yuk}.
The values $\lambda_{\rm yuk}=0.1$, $0.03$ and $0$ correspond to fermion masses $\sim 17$, $5.1$,  and $0$ GeV respectively, and we see that the temperature is largely independent of the choice of mass. This suggests that most of the energy transfer comes down through the gauge field coupling, rather than directly through the coupling to the Higgs. This conclusion may not hold for the top quark mass, which is high above the effective temperature. The dependence on the Wilson term coefficient was found to be comparable (not shown).

%Results of simulations, with or without the Wilson term, are shown in the Fig. \ref{fig:rw}.
%Differences in the temperature and chemical potential are small, comparatively the difference of the fermion energy is enormous in the Fig. \ref{fig:rw_en}.
%\begin{figure}[ht]
%\begin{center}
%\includegraphics[width=1.0\textwidth]{PIC/rw_tem_che}
%\caption{The temperature (Left plot) and chemical potential (Right plot) come from $r_w=0$ (Red line) and $r_w=0.5$ (Black line). Nq=2400, N=32, $\lambda_{\rm yuk}%$=0.03 and amH=0.42.}
%\label{fig:rw}
%\end{center}
%\end{figure}

%%%%%%%%%%%%%%%%%%%%%%%%%%%%%%%%%%%%%%%%%%%%%%%%%%%%%
\section{Fermion temperature in Cold Electroweak Baryogenesis}
\label{sec:fermT}
%%%%%%%%%%%%%%%%%%%%%%%%%%%%%%%%%%%%%%%%%%%%%%%%%%%%%%

\begin{figure}
\begin{center}
\includegraphics[width=1.0\textwidth]{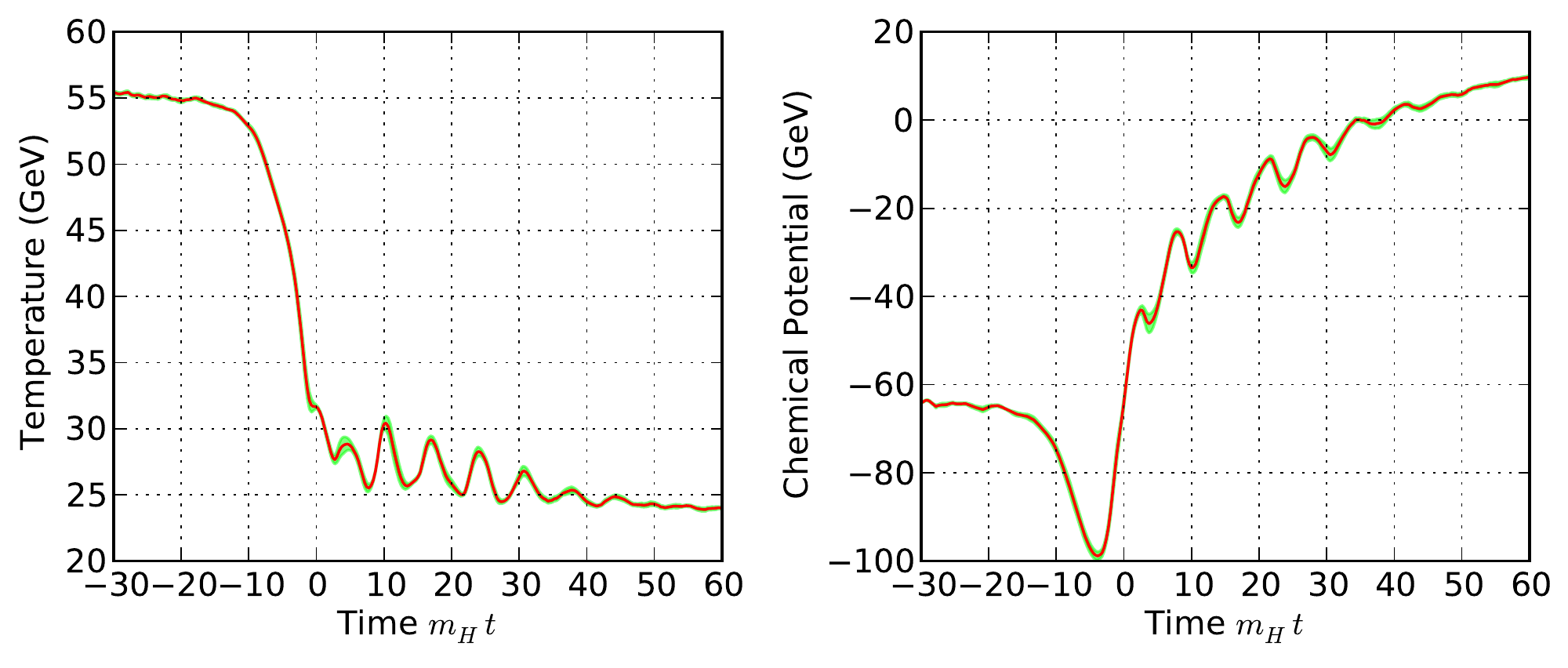}
\caption{The effective temperature and chemical potential for a $\tau_Q=30/m_H$ quench. Green bands denote statistical errors (1$\sigma$), over 8 bosonic realisations.}
\label{fig:avCH}
\end{center}
\end{figure}

Cold Electroweak Baryogenesis assumes that the Universe came out of inflation around the electroweak scale, triggering a fast electroweak quench. The out-of-equilibrium dynamics of this tachyonic transition is then responsible for Baryogenesis, without a need for a first order phase transition with bubble nucleation \cite{CEB1,CEB2,CEB3,CEB4}. It is known that the observed asymmetry can be generated in the presence of a generic additional source of CP-violation \cite{kappadep,konstandin}. But extensive work has also gone into the possibility that the CP-violation already present in the Standard Model through the complex phase in the CKM matrix may be sufficient \cite{konstandin2, salcedo1,salcedo2,CPV1,CPV2}.

The status is that a number of effective bosonic operators arise upon integrating out the fermions at finite temperature, and the coefficient functions are strongly suppressed with temperature. Bosonic simulations suggest, that the coefficient should be of order $10^{-6}$ in some normalization \cite{konstandin}, corresponding to an effective temperature in the fermions of around $1$ GeV. Although an equilibrium computation does not necessarily represent the out-of-equilibrium state during the tachyonic transition, it certainly gives the best estimate currently available. With the techniques outlined above, we are now in a situation to compute this effective temperature.

It is also known that generating the asymmetry requires the quench transition to be fairly fast \cite{quench}. This is in order for the dynamics to be violent enough that non-perturbative effects such as baryon number violation can take place. Finally, it is known that the asymmetry is created during the first or second period of the Higgs field oscillation.

We now introduce a quench time as in \cite{quench} to parametrize the flip of the mass parameters in the Higgs potential
\ba
V(\phi)=\mu^2_{\rm eff}(t)\phi^\dagger\phi +\lambda(\phi^\dagger\phi)^2,
\ea
with
\ba
\mu^2_{\rm eff}(t)&=&\mu^2\left(1-\frac{2t}{\tau_Q}\right),\qquad t<\tau_Q,\\
&=&-\mu^2,\qquad t>\tau_Q.
\ea
We then simulate the transition for different values of $\tau_Q$, in each case computing the effective temperature averaged during the first and second period of the Higgs field oscillation. We use $N_q=2400$, $N=48$, $l_{\rm yuk}=0.03$, $am_H=0.63$ and we in addition average over 8 realizations of the bosonic fields. An example of such an averaged effective temperatures  and chemical potential is shown in Fig. \ref{fig:avCH}, at a quench time of $m_H\tau_Q=30$.
%An example of such an averaged spectrum is shown in Fig. \ref{avspec}, and the effective temperatures  and chemical potential are shown in Fig. \ref{fig:avT} and Fig. \ref{fig:avCH} with 0 quench time. When $\tau_Q=30$, the effective temperature and chemical potential are given in Fig. \ref{fig:avT30} and Fig. \ref{fig:avCH30}.

%\begin{figure}
%\begin{center}
%\includegraphics[width=1.0\textwidth]{PIC/tem_sta_30qu}
%\caption{ statistics of temperature over 8 runs}
%\label{fig:avT30}
%\end{center}
%\end{figure}

%\begin{figure}
%\begin{center}
%\includegraphics[width=1.0\textwidth]{PIC/che_sta_30qu}
%\caption{ statistics of Chemical potential over 8 runs}
%\label{fig:avCH30}
%\end{center}
%\end{figure}

%The plot of the temperature of the Higgs first and second minimum and the late time limit with different quench times is given in Fig. \ref{fig:FST}.
\begin{figure}[ht]
\begin{center}
\includegraphics[width=0.6\textwidth]{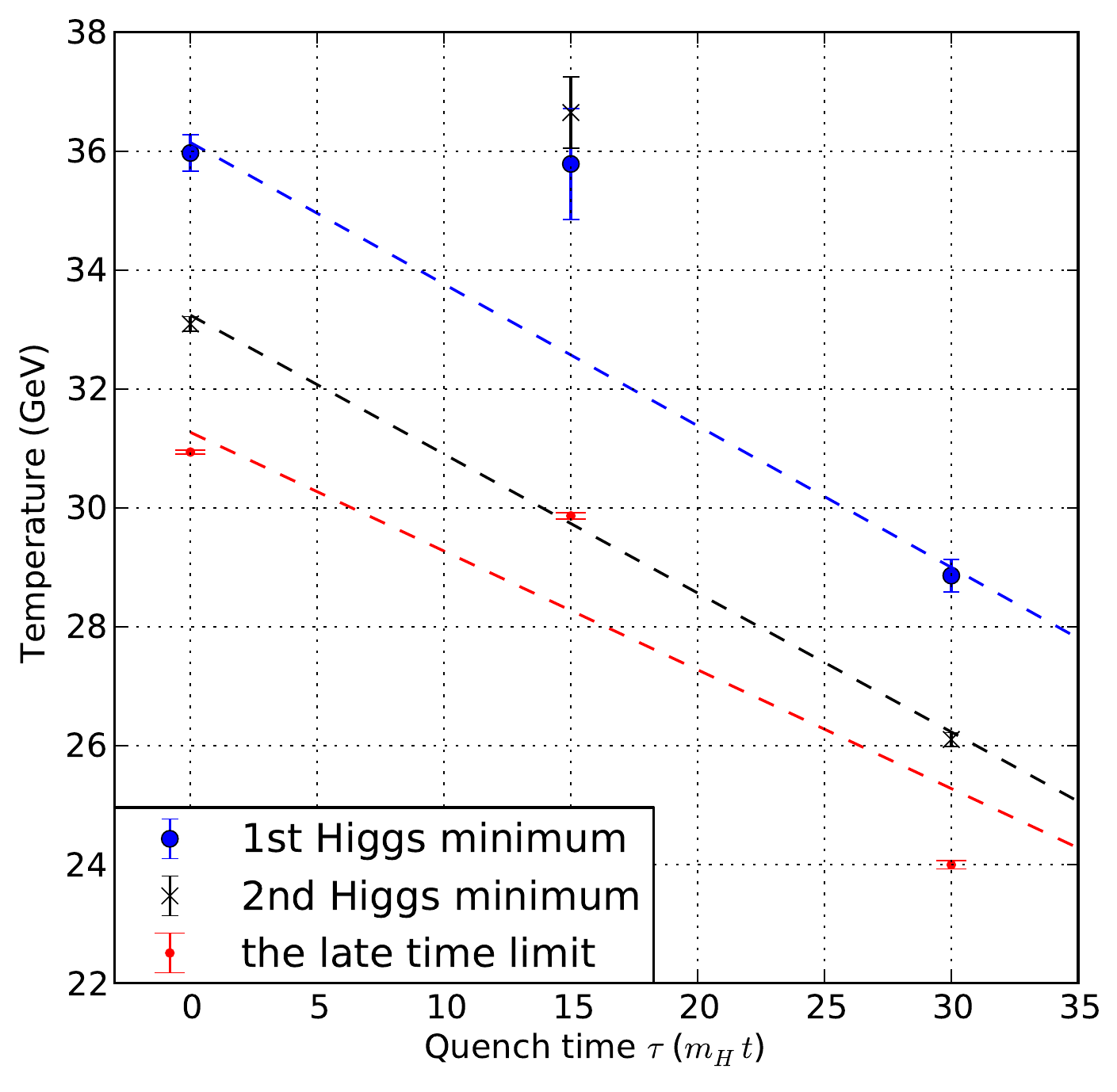}
\caption{The first and second Higgs minimum correspond to the first and second time the Higgs field rolling back to the minumum of the average Higgs field squared. Each point is the statistical result of 8 runs, with the error bar stands for $\sigma$ confidential interval. The statistical error for the late time limit is smaller than 0.2 GeV.}
\label{fig:FST}
\end{center}
\end{figure}

We finally show the effective temperatures at the first and second minimum of the Higgs field, as well as the late time limit, as a function of quench time (Fig. \ref{fig:FST}). The effective temperature oscillates in time, and the results are therefore sensitive to the exact time assigned to the Higgs minima. But the picture is clear: the temperature decreases in time, and it also decreases with quench time. For each of the measurement times (first, second minimum, late time), we can fit the temperature by a straight line.  extrapolating these to large quench times, we find that the required 1 GeV  may be reached for $m_H\tau_Q\simeq 150$. This is an order of magnitude slower than the requirements for a violent out-of-equilibrium phase during the transition \cite{quench}. We conclude that Standard Model CP-violation cannot be strong enough to generate the observed baryon asymmetry, also not in the context of Cold Electroweak Baryogenesis.

%\begin{figure}[ht]
%\begin{center}
%\includegraphics[width=1.0\textwidth]{PIC/FST2}
%\caption{The first and second Higgs minimum correspond to the first and second time the Higgs field rolling back to the minumum of the average Higgs field squared. Each point is the statistical result of 8 runs, with the error bar stands for $\sigma$ confidential interval. The statistical error for the late time limit is smaller than 0.2 GeV.}
%\label{fig:FST2}
%\end{center}
%\end{figure}

%%%%%%%%%%%%%%%%%%%%%%%%%%%%%%%%%%%%%%%%%%%%%%%%%%%%%
\section{Conclusion}
\label{sec:conc}
%%%%%%%%%%%%%%%%%%%%%%%%%%%%%%%%%%%%%%%%%%%%%%%%%%%%%%

In summary, we have carefully developed the real-time lattice implementation of the electroweak sector of the (reduced) Standard Model introduced in \cite{us2}. It includes the SU(2) gauge and Higgs fields, treated classically, and one generation of fermions, treated quantum mechanically, using the ensemble fermion method. All the fermions are taken to have the same mass, around 5 GeV. 

We found that the ensemble method converges for or order 2000 realisations of the fermions, for the dynamics and all observables except the fermion/baryon number. This is a reduction of the numerical effort by an order of magnitude, compared to what was anticipated in \cite{us2}, making the method viable for modelling the entire electroweak sector. Including the full fermion spectrum, in particular the massive top quark will make convegence slower. 
%We found that large Yukawa coupling in combination with using Wilson fermions on the lattice with a larger coefficient $r_w$ brings in additional lattice artefacts (the Wilson term dominates, a new unphysical local minimum of the energy appears). These problems will have to be dealt with on a case-by-case basis, by tuning the numerics. 
Our largest lattices ($48^3$, $N_q=2400$) fill 135GB memory, which is certainly tractable by modern supercomputers. Triple this for the full SM fermion content. 

As an application of the method, we considered the tachyonic preheating mechanism after hybrid inflation, in the context of Cold Electroweak Baryogenesis. Standard Model CP-violation gives rise to effective higher order bosonic interactions, the coefficients of which are strongly temperature dependent. In order for SM CP-violation to be sufficient for generating the observed baryon asymmetry, we need a transient stage during preheating, where the fermions have a temperature in the region of 1 GeV. We find that the transient effective temperature in the IR is 10 to 20 times higher than this, lower for slower quenches. Extrapolation to very slow quenches potentially leads to a lower preheating temperature, but we come in conflict with the need for a far-from-equilibrium state also needed for successful baryogenesis. 

We conclude that SM CP-violation is insufficient for baryogenesis, also in the Cold scenario. One caveat is that we have not included the top quark; however, since the top mass is about 5 times the largest temperature encountered here, only non-perturbative processes during the initial rolling down of the Higgs field can source it. We believe that this effect is unlikely to bridge the gap in temperature. Similarly, the reheating temperature scales with the number of relativistic degrees of freedom as $(g^*)^{-1/4}$, and so including all the SM degrees of freedom (a factor of about 4 larger compared to what we have here, depending on whether W and Z are taken to be relativistic) also does not reconcile the measured temperature with 1 GeV.

The approach to fermion dynamics considered here has now matured to the point where one may apply it to a number of phenomena, including magnetic field and gravitational wave creation during the electroweak phase transition, preheating dynamics in extensions of the SM as well as high temperature baryogenesis mechanisms where bubble walls sweep through a hot plasma. A first step is the implementation of the full three generations of the SM, with physical masses and mixings, and extensions of the Higgs sector, projects presently under way

\noindent
{\bf Acknowledgments:} The numerical simulations were implemented and performed on the COSMOS supercomputer, part of the DiRAC HPC Facility which is funded by STFC and BIS.
  ZGM wishes to thank Shuang-Yong Zhou and Paul Tognarelli for useful discussions.

%%%%%%%%%%%%%%%%%%%%%%%%%%%%%%%%%%%%%%%%%%%%%%%%%%%%%
\appendix
\section{Fermion doublers and the Wilson term}
\label{app:wilson}
%%%%%%%%%%%%%%%%%%%%%%%%%%%%%%%%%%%%%%%%%%%%%%%%%%%%%

In our simulation, we include only the spatial Wilson term to reduce the effect of fermion doublers, 
\ba
S_W&=&\int\; d^4x\;\frac{r_wa}{2}\left[ \bar\Psi D^i D_i \Psi
                          +\bar\chi D^i \del_i \chi+\bar\xi D^i \del_i \xi\right].
\ea

The temporal doubler is suppressed by choosing the initial condition carefully, so that the doubler starts out un-excited. For a long enough simulation, the doubler mode will return, but we find that this happens on a timescale much longer than the simulations presented here. 

Considering a single mode of the $U$ part (with $U$ one of the eigenspinors), which follows the difference equation 
\ba
\gamma^\mu \tilde\Delta_\mu U_k(t) + m_kU_k(t)=0,
\ea
where the $t$ is discretized into integer values, and on the first step $U_k(1)=U_ke^{i\underline k.\underline x}$.
The eigenspinor $U_k$ is the solution of
\ba
-i\gamma^0\sin k_0 U_k+ i\gamma^i\sin k_i U_k + m_k U_k =0,
\ea
where the dispersion relation  reads $\sin^2k_0 = \sum_i \sin^2k_i+m^2_k$.
So here the recurrence relation is
\ba
U_k(t+1) - U_k(t-1) = -i2\sin k_0 U_k(t).
\ea
The general solution is
\ba
U_k(t+1) = \frac{e^{-ik_0t}}{2\cos k_0}[U_k(1)e^{ik_0}+U_k(2)] + (-)^t \frac{e^{ik_0t}}{2\cos k_0}[U_k(1)e^{-ik_0}-U_k(2)].
\ea
The second term on the right-hand side is the doubler and will switch the sign from step to step.
To remove it, one needs to select the initial condition that $U_k(2)=e^{-ik_0}U_k(1)$,
and similarly, $V_k(2)=e^{ik_0}V_k(1)$ for the $V$ part.

With initial conditions chosen carefully for each mode, the temporal doubler will not be excited in the beginning.
During the simulation, we keep track of the physical and doubler parts by measuring three closest time steps,
\ba
\psi_{p}(\underline x,t) = \frac{\psi(\underline x,t+1)+2\psi(t)+\psi(\underline x,t-1)}{4},\\
\psi_{d}(\underline x,t) = \frac{\psi(\underline x,t+1)-2\psi(t)+\psi(\underline x,t-1)}{4}.
\ea
If there are only physical modes,
\ba
\label{phy}
\frac{1}{2}\langle\psi^\dagger_{pG}(\underline x,t)\psi_{pG}(\underline x,t)\rangle_e =  1,~~~~~~
\frac{1}{2}\langle\psi^\dagger_{dG}(\underline x,t)\psi_{dG}(\underline x,t)\rangle_e =  0,
\ea
and on the contrary, with only doubling modes,
\ba
\frac{1}{2}\langle\psi^\dagger_{pG}(\underline x,t)\psi_{pG}(\underline x,t)\rangle_e =  0,~~~~~~
\frac{1}{2}\langle\psi^\dagger_{dG}(\underline x,t)\psi_{dG}(\underline x,t)\rangle_e =  1.
\ea
In practical simulations, \ref{phy} is accurate up to the order of $10^{-5}$.

%%%%%%%%%%%%%%%%%%%%%%%%%%%%%%%%%%%%%%%%%%%%%%%%%%%%%
\section{Two-point functions on the lattice}
\label{sec:a1}
%%%%%%%%%%%%%%%%%%%%%%%%%%%%%%%%%%%%%%%%%%%%%%%%%%%%%

For the Dirac field, the free two-point function has the form:
\ba
G_0(x,y)  
= \int_{-\pi}^{\pi}  \frac{d^4p}{(2\pi)^4} e^{ip(x-y)}G_0(p)
=\int_{-\pi}^{\pi}  \frac{d^4p}{(2\pi)^4} \frac{e^{ip(x-y)}}{ i\gamma^\mu\sin p_\mu + m_p}.
\ea
The wave number $p$ is continuous if the size of the lattice $N$ is infinite.
For finite $N$, the integral should be substituted by a sum over $p=(2i-N+1)\pi/2$, where $i=0, 1, ... N-1$.
For any case, $p$ is periodic with period $2\pi$.

The above integral with the Minkowski signature contains on-shell poles for the real-time simulation.
Different choices of open contour integration gives the definition of Feynman, advanced or retarded Green functions, and closed contour integrals will give the density function and statistical propagator.

The full two-point function can be expanded perturbatively if the interaction of the fermion and the background is weak.
For the Higgs field background, if we choose the unitary gauge fixing and only consider zero component, the expansion is,
\ba
G(x,y)  
&=& G_0(x,y) -
\lambda_{\rm yuk}  \int_{-\pi}^{\pi} \frac{d^4k}{(2\pi)^4} e^{ikx} \phi_0(k) \int \frac{d^4p}{(2\pi)^4}e^{ip(x-y)}
G_0(p+k)G_0(p),
\label{green:higgs}
\ea
up to the first order of $\lambda_{\rm yuk} \phi_0$.

For the gauge field background,  the perturbation is 
\ba
G(x,y)  &=&G_0(x,y)+i\sum_\mu\int_{-\pi}^{\pi}\frac{d^4p}{(2\pi)^4} \frac{d^4k}{(2\pi)^4} A_\mu(k) e^{ikx}e^{ip(x-y)}
G_0(p+k)\Gamma^\mu(p+k,p) G_0(p)
\nonumber \\
& &+... 
\label{green:gauge}
\ea
where we have interpreted the gauge link as $[U_\mu(x)-1] = -iA_\mu(x)$, $[U^\dagger_\mu(x)-1] = iA_\mu(x)$ to the first order.
So the interaction vertex induced on the lattice is,
\ba
\Gamma^\mu(p+k,p)& =& 
\frac{\gamma^\mu}{2}[ e^{-i(p_\mu+k_\mu)} + e^{ip_\mu} ]
+\frac{r_w}{2} [e^{-i(p_\mu+k_\mu)} - e^{ip_\mu}].
\ea
The above perturbation respects the gauge symmetry, in the sense that the Ward identity is fulfilled in the form
\ba
i\sum_\mu (1-e^{ik_\mu}) \Gamma^\mu(p+k,p) = -i G^{-1}_0(p+k) + i G^{-1}_0(p).
\ea

%%%%%%%%%%%%%%%%%%%%%%%%%%%%%%%%%%%%%%%%%%%%%%%%%%%%%
\section{Counterterms}
\label{app:counter}
%%%%%%%%%%%%%%%%%%%%%%%%%%%%%%%%%%%%%%%%%%%%%%%%%%%%%

We introduce counterterms for the back reaction of the quantum fermions onto the classical bosonic fields. 
The full action $S_{C}$ is chosen to be,
\ba
S_{C} = -\int d^4x ~\left[\frac{(Z_3-1)}{4} W^a_{\mu \nu}W^{a,\mu \nu} + (Z_{\phi}-1) D_\mu \phi^\dagger D^{\mu} \phi + \delta V(\phi) \right],
\ea
The first term describes the screening effect when $Z_3<1$,  and will reduce the coupling constant. % Other terms will modify the interaction among background fields.
In the equation of the motion for the Higgs field, the cancellation is
\ba
(Z_\phi -1)\partial_\mu \partial^\mu \phi_0 -\frac{1}{2}\delta V'(\phi_0) 
&=&\lambda_{\rm yuk} \left[ \frac{\overline{\Psi}_u\chi +\overline{\chi}\Psi_u +\overline{\Psi}_d\xi +\overline{\xi}\Psi_d }{2}\right]
\nonumber \\
&=& i \lambda_{\rm yuk}\Tr [G(x,x)-G'(x,x)],
\label{equ:cphi}
\ea
where $G(x,y)$ and $G'(x,y)$  are Green functions for different mass eigenstates in the weak theory. 
For the gauge field
\ba
& &\frac{4(Z_3-1)}{g^2} \left(E_{n}^a(x)-E_{n}^a(x-0\right)-\frac{4(Z_3-1)}{g^2}\sum_m D_m^{ab'}\Tr \left[i\sigma^b U_{x,m}U_{x+m,n}U^\dagger_{x+n,m}U^\dagger_{x,n}\right]\nonumber\\
&=&\left[\bar\Psi_x\gamma^n i\sigma^aU_{x,n}\Psi_{x+n}+\bar\Psi_{x+n}\gamma^nU_{x,n}^\dagger i\sigma^a\Psi_x\right]-r_w\left[\bar{\Psi}_x i\sigma^a U_{x,n}\Psi_{x+n}-\bar{\Psi}_{x+n}U_{x,n}^\dagger i\sigma^a \Psi_x\right]
\nonumber \\
&=&i\Tr [i\sigma^aU_n(x)G(x+n,x) \gamma^n]+i\Tr [U^\dagger_n(x)i\sigma^aG(x,x+n) \gamma^n]
 -ir_w\Tr [i\sigma^aU_n(x)G(x+n,x)]
\nonumber \\
& &+ir_w\Tr [U^\dagger_n(x)i\sigma^aG(x,x+n)].
\label{equ:cgau}
\ea
Using the perturbative Green function \ref{green:higgs} and \ref{green:gauge}, the coefficients of the counterterms can be computed.
We perform the contour integral in continuous energy first. For instance, 
\ba
\int_{-\pi}^{\pi}\frac{d\omega}{2\pi}\frac{B}{\sin^2\omega-\sin^2c+ i\epsilon}
=-2i\frac{B}{\sin 2c},~~~~~~~~~~c \in [0,\frac{\pi}{2}].
\ea
This leaves three dimensional discretised lattice sums, which are computed numerically.
We choose the counterterm for the potential
\ba
\delta V = \frac{ct_1}{2}\phi^2 + \frac{ct_2}{4}\phi^4,
\ea
where $ct_1$ depends on the lattice spacing quadratically, and $ct_2$ depends on the logarithm of the lattice spacing.
$\phi$ can be set to be constant in \ref{green:higgs} and \ref{equ:cphi} to get the $ct_1$ and $ct_2$ quickly.
To obtain $Z_3$ and $Z_{\phi}$, one may select the field to have a particular momentum to simplify the calculation.
With our choice of lattice parameters, $0.98<Z_3<0.99$, and $Z_{\phi}\sim 1$.


\begin{thebibliography}{*}

%\cite{Saffin:2011kn}
\bibitem{us2}
  P.~M.~Saffin and A.~Tranberg,
  %``Dynamical simulations of electroweak baryogenesis with fermions,''
  JHEP {\bf 1202} (2012) 102
  [arXiv:1111.7136 [hep-ph]].
  %%CITATION = ARXIV:1111.7136;%%
  %3 citations counted in INSPIRE as of 28 Apr 2013

%\cite{Brauner:2011vb}
\bibitem{CPV1}
  T.~Brauner, O.~Taanila, A.~Tranberg and A.~Vuorinen,
  %``Temperature Dependence of Standard Model CP Violation,''
  Phys.\ Rev.\ Lett.\  {\bf 108} (2012) 041601
  [arXiv:1110.6818 [hep-ph]].
  %%CITATION = ARXIV:1110.6818;%%
  %5 citations counted in INSPIRE as of 28 Apr 2013




%\cite{D'Onofrio:2012jk}
\bibitem{sphaleron}
  M.~D'Onofrio, K.~Rummukainen and A.~Tranberg,
  %``The Sphaleron Rate through the Electroweak Cross-over,''
  JHEP {\bf 1208} (2012) 123
  [arXiv:1207.0685 [hep-ph]].
  %%CITATION = ARXIV:1207.0685;%%
  %6 citations counted in INSPIRE as of 28 Apr 2013

%\cite{Kajantie:1996mn}
\bibitem{rummukainen}
  K.~Kajantie, M.~Laine, K.~Rummukainen and M.~E.~Shaposhnikov,
  %``Is there a hot electroweak phase transition at m(H) larger or equal to m(W)?,''
  Phys.\ Rev.\ Lett.\  {\bf 77} (1996) 2887
  [hep-ph/9605288].
  %%CITATION = HEP-PH/9605288;%%
  %312 citations counted in INSPIRE as of 28 Apr 2013
  
  %\cite{Copeland:2002ku}
\bibitem{clas1}
  E.~J.~Copeland, S.~Pascoli and A.~Rajantie,
  %``Dynamics of tachyonic preheating after hybrid inflation,''
  Phys.\ Rev.\ D {\bf 65} (2002) 103517
  [hep-ph/0202031].
  %%CITATION = HEP-PH/0202031;%%
  %73 citations counted in INSPIRE as of 28 Apr 2013
  
  %\cite{Rajantie:2000nj}
\bibitem{clas2}
  A.~Rajantie, P.~M.~Saffin and E.~J.~Copeland,
  %``Electroweak preheating on a lattice,''
  Phys.\ Rev.\ D {\bf 63} (2001) 123512
  [hep-ph/0012097].
  %%CITATION = HEP-PH/0012097;%%
  %46 citations counted in INSPIRE as of 28 Apr 2013
  
  %\cite{GarciaBellido:2003wd}
\bibitem{clas3}
  J.~Garcia-Bellido, M.~Garcia-Perez and A.~Gonzalez-Arroyo,
  %``Chern-Simons production during preheating in hybrid inflation models,''
  Phys.\ Rev.\ D {\bf 69} (2004) 023504
  [hep-ph/0304285].
  %%CITATION = HEP-PH/0304285;%%
  %60 citations counted in INSPIRE as of 28 Apr 2013
  
  %\cite{DiazGil:2007dy}
\bibitem{clas4}
  A.~Diaz-Gil, J.~Garcia-Bellido, M.~Garcia Perez and A.~Gonzalez-Arroyo,
  %``Magnetic field production during preheating at the electroweak scale,''
  Phys.\ Rev.\ Lett.\  {\bf 100} (2008) 241301
  [arXiv:0712.4263 [hep-ph]].
  %%CITATION = ARXIV:0712.4263;%%
  %54 citations counted in INSPIRE as of 28 Apr 2013
  
  
  %\cite{Tranberg:2003gi}
\bibitem{CEB4}
  A.~Tranberg and J.~Smit,
  %``Baryon asymmetry from electroweak tachyonic preheating,''
  JHEP {\bf 0311} (2003) 016
  [hep-ph/0310342].
  %%CITATION = HEP-PH/0310342;%%
  %63 citations counted in INSPIRE as of 28 Apr 2013

%\cite{Tranberg:2006ip}
\bibitem{kappadep}
  A.~Tranberg and J.~Smit,
  %``Simulations of cold electroweak baryogenesis: Dependence on Higgs mass and strength of CP-violation,''
  JHEP {\bf 0608} (2006) 012
  [hep-ph/0604263].
  %%CITATION = HEP-PH/0604263;%%
  %19 citations counted in INSPIRE as of 28 Apr 2013 
%\cite{Tranberg:2009de}

\bibitem{konstandin}
  A.~Tranberg, A.~Hernandez, T.~Konstandin and M.~G.~Schmidt,
  %``Cold electroweak baryogenesis with Standard Model CP violation,''
  Phys.\ Lett.\ B {\bf 690} (2010) 207
  [arXiv:0909.4199 [hep-ph]].
  %%CITATION = ARXIV:0909.4199;%%
  %23 citations counted in INSPIRE as of 28 Apr 2013

%\cite{Tranberg:2006dg}
\bibitem{quench}
  A.~Tranberg, J.~Smit and M.~Hindmarsh,
  %``Simulations of cold electroweak baryogenesis: Finite time quenches,''
  JHEP {\bf 0701} (2007) 034
  [hep-ph/0610096].
  %%CITATION = HEP-PH/0610096;%%
  %20 citations counted in INSPIRE as of 28 Apr 2013

  
  %\cite{Aarts:1999zn}
\bibitem{ferm1}
  G.~Aarts and J.~Smit,
  %``Particle production and effective thermalization in inhomogeneous mean field theory,''
  Phys.\ Rev.\ D {\bf 61} (2000) 025002
  [hep-ph/9906538].
  %%CITATION = HEP-PH/9906538;%%
  %51 citations counted in INSPIRE as of 28 Apr 2013
  
  %\cite{Aarts:1998td}
\bibitem{ferm2}
  G.~Aarts and J.~Smit,
  %``Real time dynamics with fermions on a lattice,''
  Nucl.\ Phys.\ B {\bf 555} (1999) 355
  [hep-ph/9812413].
  %%CITATION = HEP-PH/9812413;%%
  %39 citations counted in INSPIRE as of 28 Apr 2013
  
  %\cite{Borsanyi:2008eu}
\bibitem{ferm3}
  S.~Borsanyi and M.~Hindmarsh,
  %``Low-cost fermions in classical field simulations,''
  Phys.\ Rev.\ D {\bf 79} (2009) 065010
  [arXiv:0809.4711 [hep-ph]].
  %%CITATION = ARXIV:0809.4711;%%
  %12 citations counted in INSPIRE as of 28 Apr 2013
  

%\cite{Hebenstreit:2013qxa}
\bibitem{ferm4}
  F.~Hebenstreit, Jür.~Berges and D.~Gelfand,
  %``Simulating fermion production in 1+1 dimensional QED,''
  Phys.\ Rev.\ D {\bf 87} (2013) 105006
  [arXiv:1302.5537 [hep-ph]].
  %%CITATION = ARXIV:1302.5537;%%
  %2 citations counted in INSPIRE as of 30 Jul 2013

%\cite{Hebenstreit:2013baa}
\bibitem{ferm5}
  F.~Hebenstreit, Jür.~Berges and D.~Gelfand,
  %``Real-time dynamics of string breaking,''
  arXiv:1307.4619 [hep-ph].
  %%CITATION = ARXIV:1307.4619;%%
  
%\cite{Saffin:2011kc}
\bibitem{us1}
  P.~M.~Saffin and A.~Tranberg,
  %``Real-time Fermions for Baryogenesis Simulations,''
  JHEP {\bf 1107} (2011) 066
  [arXiv:1105.5546 [hep-ph]].
  %%CITATION = ARXIV:1105.5546;%%
  %5 citations counted in INSPIRE as of 28 Apr 2013
  
    
%\cite{Brauner:2012gu}
\bibitem{CPV2}
  T.~Brauner, O.~Taanila, A.~Tranberg and A.~Vuorinen,
  %``Computing the temperature dependence of effective CP violation in the standard model,''
  JHEP {\bf 1211} (2012) 076
  [arXiv:1208.5609 [hep-ph]].
  %%CITATION = ARXIV:1208.5609;%%
  %1 citations counted in INSPIRE as of 28 Apr 2013

%\cite{GarciaBellido:1999sv}
\bibitem{CEB1}
  J.~Garcia-Bellido, D.~Y.~Grigoriev, A.~Kusenko and M.~E.~Shaposhnikov,
  %``Nonequilibrium electroweak baryogenesis from preheating after inflation,''
  Phys.\ Rev.\ D {\bf 60} (1999) 123504
  [hep-ph/9902449].
  %%CITATION = HEP-PH/9902449;%%
  %152 citations counted in INSPIRE as of 28 Apr 2013
  
%\cite{Krauss:1999ng}
\bibitem{CEB2}
  L.~M.~Krauss and M.~Trodden,
  %``Baryogenesis below the electroweak scale,''
  Phys.\ Rev.\ Lett.\  {\bf 83} (1999) 1502
  [hep-ph/9902420].
  %%CITATION = HEP-PH/9902420;%%
  %91 citations counted in INSPIRE as of 28 Apr 2013
  
%\cite{Copeland:2001qw}
\bibitem{CEB3}
  E.~J.~Copeland, D.~Lyth, A.~Rajantie and M.~Trodden,
  %``Hybrid inflation and baryogenesis at the TeV scale,''
  Phys.\ Rev.\ D {\bf 64} (2001) 043506
  [hep-ph/0103231].
  %%CITATION = HEP-PH/0103231;%%
  %58 citations counted in INSPIRE as of 28 Apr 2013
  

%\cite{Hernandez:2008db}
\bibitem{konstandin2}
  A.~Hernandez, T.~Konstandin and M.~G.~Schmidt,
  %``Sizable CP Violation in the Bosonized Standard Model,''
  Nucl.\ Phys.\ B {\bf 812} (2009) 290
  [arXiv:0810.4092 [hep-ph]].
  %%CITATION = ARXIV:0810.4092;%%
  %19 citations counted in INSPIRE as of 28 Apr 2013

%\cite{GarciaRecio:2009zp}
\bibitem{salcedo1}
  C.~Garcia-Recio and L.~L.~Salcedo,
  %``CP violation in the effective action of the Standard Model,''
  JHEP {\bf 0907} (2009) 015
  [arXiv:0903.5494 [hep-ph]].
  %%CITATION = ARXIV:0903.5494;%%
  %9 citations counted in INSPIRE as of 28 Apr 2013
  
  %\cite{Salcedo:2011hy}
\bibitem{salcedo2}
  L.~L.~Salcedo,
  %``Leading order one-loop CP and P violating effective action in the Standard Model,''
  Phys.\ Lett.\ B {\bf 700} (2011) 331
  [arXiv:1102.2400 [hep-ph]].
  %%CITATION = ARXIV:1102.2400;%%
  %5 citations counted in INSPIRE as of 28 Apr 2013
  
\end{thebibliography}
\end{document}